\documentclass[preprint,preprintnumbers,nofootinbib]{revtex4-1}

\AtBeginDocument{%
  \newwrite\bibnotes
  \def\bibnotesext{Notes.bib}
  \immediate\openout\bibnotes=\jobname\bibnotesext
  \immediate\write\bibnotes{@CONTROL{REVTEX41Control}}
  \immediate\write\bibnotes{@CONTROL{%
    apsrev41Control,author="08",editor="1",pages="1",title="0",year="1"}}
  \if@filesw
    \immediate\write\@auxout{\string\citation{apsrev41Control}}%
  \fi
}
\usepackage[normalem]{ulem}
\usepackage{graphicx}
\usepackage{dcolumn}
\usepackage{bm}
\usepackage{amsmath}
\usepackage{amsfonts} 
\usepackage{latexsym}
\usepackage{bbm}
\usepackage{color}
\usepackage{amssymb}
\usepackage{amsthm}
\usepackage{epsfig}
\usepackage{physics}
\usepackage{comment}
\usepackage{tikz-feynhand}
\usepackage{subcaption}
\usepackage{booktabs}
\usepackage{ragged2e}
\usepackage{tabularx}
\usepackage{hyperref}
\hypersetup{
setpagesize=false,
 bookmarksnumbered=true,
 bookmarksopen=true,
 colorlinks=true,
 linkcolor=blue,
 citecolor=red,
}

\begin{document}

\title{Constraints on the Real Singlet Scalar Dark Matter Model from Higgs Self-Couplings}

\author{Hao-Yu Luo}
\author{Kun-Peng Chen}
\author{Han Hu}
\author{Yun Wei}
\author{Ming Zhong}
\email{zhongm@nudt.edu.cn}
\affiliation{College of Science, National University Of Defense Technology, Changsha, China}

\vspace{2cm}

\begin{abstract}
Higgs portal models constitute one of the theoretically best-motivated classes of Standard Model extensions. Portal-sector particles generically induce deviations of Higgs self-couplings, and precision measurements of these couplings can accordingly be leveraged to constrain the physical properties of the portal sector. Taking the real singlet scalar dark matter model as a representative test case, the present study examines the constraints imposed on the model's parameter space by Higgs trilinear coupling measurements and perturbative unitarity requirements. For the low dark matter mass regime, the constraining strength delivered by the current highest-precision $\kappa_3$ measurement is notably inferior to the exclusion limit derived from perturbative unitarity requirements, while the perturbative unitarity bound itself remains considerably less restrictive than the constraints extracted from the extant experimental upper limit on the Higgs invisible decay branching ratio, $\mathrm{BR}_{\mathrm{inv}} \leq 0.107$. Both the coupling modifiers $\kappa_3$ and $\kappa_4$ exhibit a well-defined decoupling behavior, with their respective magnitudes monotonically decreasing as the dark matter mass $m_S$ increases. As a direct consequence, the corrections to both couplings asymptotically approach zero for dark matter masses in excess of 60 TeV, when the parameter $a_2$ remains within its perturbative validity domain. Neither existing collider measurements of the Higgs trilinear coupling nor measurements targeted for planned future collider facilities possess the requisite sensitivity to impose more stringent constraints on the model’s viable parameter space than the well-established bounds derived from experimental probes of dark matter relic density, invisible Higgs decay signatures, dark matter direct detection, and dark matter indirect detection. Nevertheless, under the scenario that the experimental measurement uncertainty of $\kappa_3$ reaches the $\mathcal{O}(0.1\%)$ scale or lower, non-negligible constraints acting on a subset of the large mass parameter space associated with the weakly interacting massive particle (WIMP) dark matter scenario can be expected.
\end{abstract}
\maketitle

%%%%%%%%%%%%%%%%%%%%%%%%%%%%%%%%%%%%%%%%%%%%%%%%
%		          
%%%%%%%%%%%%%%%%%%%%%%%%%%%%%%%%%%%%%%%%%%%%%%%%

\section{Introduction}
\label{sec:intro}

The evidence for the existence of dark matter (DM) constitutes a profound challenge to the Standard Model (SM) of particle physics. To account for it, numerous new physics models have been proposed, among which the Higgs portal models are recognized as one of the most theoretically well-motivated classes of SM extensions. These models postulate that DM interacts with the SM sector via couplings to the Higgs boson. Accordingly, the production, annihilation, and scattering processes of DM off ordinary matter occur in manners that are amenable to experimental verification.

The phenomenology of Higgs portal DM models has been comprehensively investigated \cite{Arcadi:2019lka,Lebedev:2021xey}. The observable phenomena are mainly governed by the DM mass and the Higgs portal couplings. Current leading constraints on these parameters are derived from relic density measurements \cite{Planck:2018vyg}, direct detection experiments \cite{aprile2025wimpdarkmattersearch, LZ:2024zvo}, indirect detection searches \cite{Ackermann_2015, Alemanno_2022}, and collider probes of invisible Higgs decays \cite{ATLAS:2023tkt}. While the interplay among these constraints delineates some constrained viable parameter spaces, the precise nature of DM and the detailed structure of the Higgs portal coupling remain unresolved. This ongoing uncertainty motivates continuous theoretical refinements and the development of next generation experimental designs aimed at deciphering this fundamental enigma.

An implication of Higgs portal DM is that it induces modifications to the triple Higgs coupling $\lambda_{hhh}$ and the quartic Higgs coupling $\lambda_{hhhh}$. The magnitude of these modifications is intrinsically correlated with the strength of the portal coupling and the mass of the DM candidate. Accordingly, precision measurements of these Higgs self-couplings can serve as indirect probes for DM, and in turn provide a valuable complement to collider-based probes of invisible Higgs decay which are kinematically inaccessible in the large DM mass regime.

Precision measurements of the Higgs sector have emerged as a critical avenue for probing physics beyond the SM \cite{Braathen_2020, Braathen_2019, Kanemura_2017, Bahl_2022}. As fundamental parameters characterizing the structure of the Higgs potential, the triple and quartic Higgs couplings play a pivotal role in governing the dynamics of electroweak phase transitions. Recent combined results from the ATLAS and CMS collaborations constrain the coupling modifier $\kappa_3 = \lambda_{hhh}/\lambda_{hhh}^{(\mathrm{SM})}$ to the interval $[-0.71, 6.1]$ \cite{CMS:2026nuu}, while perturbative unitarity imposes theoretical constraints on $\kappa_3$ and $\kappa_4 = \lambda_{hhhh}/\lambda_{hhhh}^{(\mathrm{SM})}$ \cite{Stylianou_2024, Liu_2018, Chang_2020}. Future collider facilities are projected to substantially tighten constraints on $\kappa_3$ to a precision of a few percent, and are expected to yield the experimental bounds on $\kappa_4$ \cite{fujii2015physicscaseinternationallinear, fujii2018physicscase250gev, Roloff_2020, cepeda2019higgsphysicshllhchelhc, Gon_alves_2018, contino2016physics100tevpp}.

In this work, we employ Higgs self-coupling measurements to constrain the parameter space of the real singlet scalar DM model. This model is the simplest Higgs-portal extension of the SM, constructed by introducing a real singlet scalar field that preserves a $\mathrm{Z}_2$ symmetry \cite{SILVEIRA1985136}. We focus on examining the validity of the existing viable parameter regions of this model using both the latest available and future projected Higgs self-coupling measurements. Specifically, in the low-mass dark matter regime, we compare the parameter constraints derived from measurements of $\kappa_3$ and Higgs invisible decays; in the high-mass dark matter regime where Higgs invisible decays are kinematically forbidden, we treat the measurement of $\kappa_3$ as an independent constraint on the model, and analyze its impact on the parameter space constraint. Our analysis demonstrates that existing collider measurements and currently planned future collider measurements are fully consistent with all viable parameter regions obtained from existing observations of DM relic density, invisible Higgs decays, direct detection, and indirect detection. Nontrivial constraints may be imposed on a subset of the parameter space associated with large-mass weakly interacting massive particle (WIMP) dark matter candidates, under the extreme precision benchmark regime defined by a $\kappa_3$ measurement uncertainty of order $\mathcal{O}(0.1\%)$. Such a precision level is virtually unattainable in experimental practice. Moreover, when $\kappa_3$ lies within the interval $(0.9999, 1.0001)$, the constraint it imposes on the model parameter space is comparable to that derived from the current experimental result $\mathrm{BR}_{\mathrm{inv}} \leq 0.107$ for the invisible decay branching ratio of the Higgs boson. This requires the measurement uncertainty of $\kappa_3$ to be lower than one ten-thousandth, which is a fortiori unattainable experimentally. Therefore, the low-mass parameter region constrained by Higgs invisible decay remains completely unaffected.

Section~\ref{sec:model} provides a brief introduction to the $\mathrm{Z}_2$ extended standard model ($\mathrm{Z}_2\mathrm{xSM}$) with a singlet scalar DM. In Section~\ref{sec:phenomenology}, we elaborate on the phenomenological properties of DM in this model, and summarize the viable parameter regions for WIMP, SIMP, and freeze-in at strong coupling dark matter candidates. The effective trilinear and quartic Higgs couplings of the model are analyzed in Section~\ref{sec:kappa}. We adopt the one-loop corrected Higgs potential to define the Higgs trilinear and quartic couplings, as well as their corresponding coupling modifiers $\kappa_3$ and $\kappa_4$. In Section~\ref{sec:constraints}, we examine the consistency between the parameter regions outlined in Section~\ref{sec:phenomenology} and existing $\kappa_3$ measurements, as well as the perturbative unitarity constraint. We further discuss the potential constraints that can be obtained from future measurements at collider facilities. Finally, we conclude the present work in Section~\ref{sec:conclusion}.

%%%%%%%%%%%%%%%%%%%%%%%%%%%%%%%%%%%%%%%%%%%%%%%%
%		        
%%%%%%%%%%%%%%%%%%%%%%%%%%%%%%%%%%%%%%%%%%%%%%%%

\section{The $\mathbf{Z}_2\mathbf{xSM}$ Model}
\label{sec:model}

The minimal Higgs portal model is constructed by extending the SM with an additional real scalar field $S$, which is a singlet under the SM gauge groups and obeys a discrete $\mathrm{Z}_2$ symmetry satisfying $S \rightarrow -S$ \cite{SILVEIRA1985136,Cheung_2012, Cline_2013, Feng_2015, yu2025newconstraintssingletscalar, Belanger:2026ctm}. The Lagrangian of the model reads
\begin{align}
	\mathcal{L} = \frac{1}{2} \partial^\mu S\, \partial_\mu S - \frac{b_2}{2} S^2 - \frac{b_4}{4} S^4 - a_2 S^2 H^\dagger H + \mathcal{L}_{\mathrm{SM}},
\end{align}
where $H$ denotes the SM Higgs doublet.

Notwithstanding its minimal structure, this model yields rich theoretical and phenomenological implications. The general properties of this model have been extensively explored in literatures over the past decades \cite{SILVEIRA1985136,Burgess:2000yq,Gonderinger_2010,Cheung_2012,Cline_2013,Feng_2015,Han_2015,Escudero_Abenza_2025,yu2025newconstraintssingletscalar}. Recent studies have confirmed that, consistent with Higgs self-coupling measurements, this model can maintain vacuum stability up to the Planck scale \cite{Athron_2018}, and is capable of accommodating a strongly first-order electroweak phase transition that gives rise to cosmic baryon asymmetry and stochastic gravitational waves \cite{Vaskonen_2017, ghorbani2019stronglyfirstorderphasetransition,Bosse:2026bdk}. The imposed $\mathrm{Z}_2$ symmetry ensures the stability of the singlet field, enabling it to serve as a viable DM candidate via the Higgs portal coupling \cite{Gonderinger_2010}. Accordingly, experimental investigation of this model holds significant potential for addressing multiple outstanding fundamental questions in both particle physics and cosmology.

The non-zero vacuum expectation value of the Higgs field, $\left\langle H\right\rangle =\frac{1}{\sqrt{2}}(0,v)^T$, induces spontaneous electroweak symmetry breaking. The tree-level masses of the Higgs boson $h$ and the scalar particle $S$ are given by
\begin{equation}
	m_h^2= m^2 + \frac{\lambda v^2}{2}  = \frac{\lambda v^2}{3}, \qquad m_S^2 = b_2 + a_2 v^2.
\end{equation}
In addition to the full set of SM parameters, the model introduces three new independent free parameters: the Higgs portal coupling $a_2$, the scalar particle mass $m_S$, and the quartic self-coupling $b_4$. The portal coupling $a_2$ mediates all interactions between the singlet scalar and SM particles, and thus constitutes the core parameter for deriving experimental constraints on the model. The singlet self-coupling $b_4$ primarily determines the perturbativity properties and vacuum structure. Non-trivial constraints on these parameters are derived from theoretical considerations, including vacuum stability, unitarity, triviality, and perturbativity \cite{cynolter2004noteunitarityconstraintsmodel, Davoudiasl_2005, Lerner_2009, Gonderinger_2010, kang2013unitarityconstraintsstandardmodel, Lopes_2026}.

\section{Dark Matter Phenomenology of the Model}
\label{sec:phenomenology}

The DM phenomenology of this model has been extensively studied \cite{SILVEIRA1985136,Gonderinger_2010,Lerner_2009,Han_2015,Feng_2015,Cheung_2012,yu2025newconstraintssingletscalar,Cline_2013,Escudero_Abenza_2025,Belanger:2026ctm,Bernal_2019,arcadi2024higgsportaldarkmatter}. Experimental constraints on the parameter space of the model are primarily derived from relic density measurements \cite{Planck:2018vyg}, direct detection experiments \cite{aprile2025wimpdarkmattersearch,LZ:2024zvo}, indirect detection searches \cite{Ackermann_2015,Alemanno_2022}, and collider searches for invisible Higgs decays \cite{ATLAS:2023tkt}.

Direct detection experiments derive constraints on the interaction cross section between DM and nucleons by searching for rare scattering events between DM particles in the Milky Way and target nuclei. In 2025, the LUX-ZEPLIN (LZ) experiment released the largest dataset ever collected by any DM detection experiment to date. No WIMP signals were observed within the low-mass range of 3-9 GeV in the LZ dataset, and the experiment yielded world-leading constraints on the relevant parameter space \cite{LZ:2024zvo}.

Within the $\mathrm{Z}_2\mathrm{xSM}$, the scattering of DM particles off nucleons proceeds via $t$-channel Higgs exchange. The corresponding spin-independent scattering cross section at leading order (LO) is given by \cite{SILVEIRA1985136, Cheung_2012}
\begin{equation}
	\label{dd}
	\sigma_{\mathrm{SI}}^{(\mathrm{LO})} = \frac{a_2^2 f_N^2 m_N^4}{\pi (m_S + m_N)^2 m_h^4},
\end{equation}
where $f_N = 1/3$ denotes the effective form factor associated with the nucleon-Higgs coupling, and $m_N$ represents the nucleon mass.

Indirect detection searches for DM signatures by observing SM particles produced from DM annihilation or decay. In the $\mathrm{Z}_2\mathrm{xSM}$, the imposed $\mathrm{Z}_2$ symmetry guarantees the absolute stability of DM particles, hence annihilation into SM particles is the only physically relevant process for indirect detection. Multiple independent astrophysical observations have been utilized to constrain this annihilation process, including gamma-ray surveys of DM-dominated dwarf spheroidal galaxies in the vicinity of the Milky Way, measurements of spectral distortions of the cosmic microwave background, and searches for anomalous antiparticle fluxes in cosmic rays. To date, observations from the Fermi Large Area Telescope and high-energy ground-based gamma-ray observatories have placed stringent upper bounds on the DM annihilation cross section  \cite{Ackermann_2015}. 

Dark matter annihilation into gamma rays proceeds via $s$-channel Higgs exchange, i.e. $SS \rightarrow h^* \rightarrow \gamma\gamma$. The LO thermally averaged annihilation cross section is given by \cite{Profumo_2010}
\begin{equation}
	\label{idd}
	\langle \sigma v \rangle_{\gamma\gamma}^{(\mathrm{LO})}
	= a_2^2 \frac{\alpha^2}{32\pi^3}
	\frac{s}{(s-m_h^2)^2 + \Gamma_h^2 m_h^2}
	\left| \sum_i N_{ci} e_i^2 F(\tau) \right|^2,
\end{equation}
where $s\approx 4m_S^2$ denotes the squared center-of-mass energy under the nonrelativistic approximation. Here, $\alpha$ and $\Gamma_h$ correspond to the electromagnetic fine-structure constant and the total decay width of the Higgs boson, respectively. The index $i$ runs over all SM particles propagating in the interaction loop, and $N_{ci}$, $e_i$ and $F(\tau)$ denote the relevant color factor, electric charge and loop function of the $i$-th particle, respectively. The annihilation cross section receives a significant enhancement when the DM mass satisfies $m_S \approx m_h/2$, which corresponds to the Higgs resonance regime.

Collider searches for invisible Higgs decays provide complementary constraints on the low-mass region of the model. When the DM mass satisfies $m_S < m_h/2$, the Higgs boson can decay into a pair of on-shell DM particles which escape the collider detector without leaving visible energy deposits, leading to a non-zero invisible branching fraction of the Higgs boson. The LO decay width of this process is given by \cite{Cheung_2012}
\begin{equation}
	\label{hid}
	\Gamma_{\mathrm{inv}}^{(\mathrm{LO})}
	= \frac{a_2^2 v^2}{8\pi m_h}
	\operatorname{Re} \sqrt{1 - \frac{4m_S^2}{m_h^2}}.
\end{equation}

Current measurements from LHC experiments have established stringent upper bounds on the invisible decay width of the Higgs boson. For instance, the ATLAS collaboration derived the limit $\mathrm{BR}_{\mathrm{inv}} \leq 0.107$ utilizing data from LHC Run I and Run II \cite{ATLAS:2023tkt}.

The thermal relic paradigm stands as a highly compelling hypothesis for the origin of DM. Currently, the most accurate measurement of DM relic density is derived from observations of anisotropies in cosmic microwave background (CMB) radiation, which serves as a core criterion for evaluating a broad range of theoretical DM models. Leveraging the highest-precision CMB sky map acquired to date, the Planck collaboration has measured the cold DM density parameter of the Universe within the standard $\Lambda$CDM cosmological model, yielding the result \cite{Planck:2018vyg}
\begin{equation}
	\Omega_c h^2 = 0.1200 \pm 0.0012.
\end{equation}
CMB observation results from multiple independent experiments, such as SPT-3G, are also in excellent agreement with this measured value.

Within the standard cosmological framework where the Universe remains radiation-dominated over the entire epoch before Big Bang nucleosynthesis (BBN), the observed DM relic abundance can be naturally explained if DM was initially in thermal and chemical equilibrium with the SM plasma before experiencing subsequent freeze-out. The canonical realization of this paradigm is WIMP, whose relic density is governed by 2-to-2 annihilation processes into SM particles \cite{Arcadi:2024ukq}. Alternatively, if the coupling between the dark sector and the visible sector is severely suppressed, DM may never reach thermal and chemical equilibrium, and can instead be generated via the freeze-in mechanism. This is the FIMP scenario \cite{Hall:2009bx}. In scenarios with low reheating temperatures, there exists a smooth transition between these two apparently disconnected regimes \cite{Bernal_2019, arcadi2024higgsportaldarkmatter}. Another theoretically well-motivated possibility is that the DM relic density is predominantly determined by number-changing processes that occur within a secluded dark sector. In this framework, the SIMP scenario emerges when the dominant processes are 3-to-2 or 4-to-2 annihilations among DM particles. However, in the standard radiation-dominated cosmological background, the SIMP mechanism is typically subject to strict constraints. It becomes fully viable when chemical freeze-out occurs during an early matter-dominated era \cite{Belanger:2026ctm}.

We summarize several distinct and representative parameter spaces that have been explicitly identified in extensive relevant LO studies of the $\mathrm{Z}_2\mathrm{xSM}$ model, which will be tested using existing measured data and projected future data of the Higgs self-coupling in Section~\ref{sec:constraints}.
\begin{itemize}
	\item For the freeze-out mechanism of the WIMP scenario, two distinct parameter regions consistent with existing experimental data have been explicitly identified, namely the resonance mass (WIMP-RM) region and the large mass (WIMP-LM) region. In the WIMP-RM region, the DM mass satisfies $60.5\ \mathrm{GeV} < m_S < 62.5\ \mathrm{GeV}$, and the portal coupling falls in the range of $1.7\times 10^{-4} < a_2 < 4.7\times 10^{-4}$ \cite{ yu2025newconstraintssingletscalar}. The corresponding parameter space of the WIMP-LM region, which satisfies $31\ \mathrm{TeV} < m_S < 47\ \mathrm{TeV}$ and $5 < a_2 < 7.5$, is given in Ref. \cite{Escudero_Abenza_2025}.
	
	\item The SIMP mechanism generally requires sizable dark matter self-interactions and can give rise to distinctive phenomenological signatures. Nevertheless, within the standard paradigm of a radiation-dominated cosmological background, the SIMP scenario is typically subject to severe constraints. Under the assumption of an early matter-dominated Universe, the SIMP scenario allows for a broad range of DM mass spanning from $10^{-1}\ \mathrm{GeV}$ to $10^{12}\ \mathrm{GeV}$, and the portal coupling falls in the range of $10^{-5}<a_2<10$ \cite{Belanger:2026ctm}.	
	
	\item Conventional theoretical formulations of freeze-in DM production are typically predicated on a post-inflationary reheating temperature \(T_R \gtrsim m_{\rm DM}\), alongside feeble portal couplings of characteristic magnitude \(\mathcal{O}(10^{-11})\). This framework constitutes the canonical feebly interacting massive particle (FIMP) scenario, whose associated parameter space remains compatible with the vast majority of existing observational constraints. In an alternative freeze-in regime defined by \(T_R < m_{\rm DM}\), DM production rates are inherently subject to Boltzmann suppression. To reproduce the cosmologically measured DM relic abundance, the required portal coupling may reach values as large as \(\mathcal{O}(1)\) \cite{arcadi2024higgsportaldarkmatter}, a magnitude sufficiently elevated to render the scenario experimentally verifiable via direct DM detection searches and precision measurements of invisible Higgs boson decay widths. We designate this distinct coupling regime as freeze-in at strong coupling (FISC). As the portal coupling strength increases further, the coupled DM sector achieves full thermal equilibrium, and the parameter trajectory corresponding to the observed relic abundance smoothly converges to the canonical thermal freeze-out solution.
\end{itemize}

\section{The Higgs Potential and the Effective Higgs Self-Coupling}
\label{sec:kappa}

The singlet scalar can contribute to the effective potential of the Higgs field via quantum corrections, and this problem was formally investigated in Ref.~\cite{Gonderinger_2010} to examine the vacuum stability of the $\mathrm{Z}_2\mathrm{xSM}$ model. The effective potential for the Higgs field is expressed as
\begin{equation}
	\label{eq:V_eff}
	V_{\mathrm{eff}} = V_{\mathrm{tree}} + V_{\mathrm{CW}},
\end{equation}
where
\begin{equation}
	V_{\mathrm{tree}} = \frac{m^2}{2} h^2 + \frac{\lambda}{24} h^4
\end{equation}
denotes the tree-level potential, and
\begin{equation}
	V_{\mathrm{CW}} = \frac{1}{64\pi^2} \sum_j n_j m_j^4(h) \left[ \ln\frac{m_j^2(h)}{\mu^2} - c_j \right]
\end{equation}
corresponds to the one-loop Coleman-Weinberg potential derived under the $\overline{\mathrm{MS}}$ renormalization scheme. The summation is performed over all degrees of freedom of the SM as well as the additional real scalar $S$. Compared with the contribution from the top quark, the contributions from the $W$ and $Z$ bosons, Goldstone bosons, and all other fermions are negligibly small. Accordingly, only the contributions from the top quark and the scalar $S$ are retained in this derivation, yielding the approximate form
\begin{equation}
	\label{eq:V_CW}
	V_{\mathrm{CW}} \approx \frac{n_S m_S^4(h)}{64\pi^2} \left[ \ln\frac{m_S^2(h)}{\mu^2} - c_S \right] + \frac{n_t m_t^4(h)}{64\pi^2} \left[ \ln\frac{m_t^2(h)}{\mu^2} - c_t \right],
\end{equation}
where $n_S = 1$, $n_t = -12$, and $c_S = c_t = 3/2$. The field-dependent squared masses are given by
\begin{equation}
	m_S^2(h) = b_2 + a_2 h^2, \qquad m_t^2(h) = \frac{1}{4} y_t^2 h^2.
\end{equation}
For brevity, the dependence of the masses, couplings and the field $h=h(\mu)$ on the renormalization scale $\mu$ is suppressed in the above expressions. The corresponding $\beta$-functions and anomalous dimensions can be derived from the scale invariance condition of the effective potential, namely $\frac{dV_{\mathrm{eff}}}{d\mu} = 0$ \cite{Gonderinger_2010}.

The effective triple and quartic Higgs self-couplings can be obtained by computing the third- and fourth-order partial derivatives of the effective potential with respect to $h$
\begin{equation}
	\label{eq:lamuda}
	\lambda_{hhh} = \left. \frac{\partial^3 V_{\mathrm{eff}}}{\partial h^3} \right|_{h=v}, \qquad
	\lambda_{hhhh} = \left. \frac{\partial^4 V_{\mathrm{eff}}}{\partial h^4} \right|_{h=v}.
\end{equation}
Accordingly, the coupling modifiers $\kappa_3$ and $\kappa_4$ are defined as
\begin{equation}
	\kappa_3 \equiv \frac{\lambda_{hhh}}{\lambda_{hhh}^{\mathrm{SM}}}, \qquad
	\kappa_4 \equiv \frac{\lambda_{hhhh}}{\lambda_{hhhh}^{\mathrm{SM}}}.
\end{equation}
Based on Eqs.~\eqref{eq:V_eff} through \eqref{eq:lamuda}, we derive the one-loop corrected expressions for the triple and quartic Higgs self-couplings in the $\overline{\mathrm{MS}}$ renormalization scheme as follows
\begin{align}
	\lambda_{hhh}^{(1)}
	&= \frac{3m_h^2}{v}
	+ \frac{1}{64\pi^2} \left( 24 a_2^2 v \ln\frac{m_S^2}{\mu^2} + \frac{2 a_2^3 v^3}{m_S^2} \right) \nonumber \\
	&\quad - \frac{12}{64\pi^2} \left( 6 y_t^4 v \ln\frac{y_t^2 v^2}{2\mu^2} + \frac{y_t^4 v}{2} \right), \\
	\lambda_{hhhh}^{(1)}
	&= \frac{3m_h^2}{v^2}
	+ \frac{1}{64\pi^2} \left( 24 a_2^2 \ln\frac{m_S^2}{\mu^2} + \frac{96 a_2^3 v^2}{m_S^2} - \frac{32 a_2^4 v^4}{m_S^4} \right) \nonumber \\
	&\quad - \frac{12}{64\pi^2} \left( 6 y_t^4 \ln\frac{y_t^2 v^2}{2\mu^2} + 16 y_t^4 \right),
\end{align}
where $m_h^2$ denotes the tree-level squared mass of the Higgs boson. The corresponding couplings for the SM can be recovered by setting $a_2=0$.

The relation between $m_h^2$ and the curvature mass $\left[M_h^2\right]_V = \left. \frac{\partial^2 V_{\mathrm{eff}}}{\partial h^2} \right|_{h=v}$ is given by \cite{Braathen_2020}
\begin{equation}
\left[M_h^2\right]_V = m_h^2 + \left( -\frac{1}{v}\frac{\partial}{\partial h} + \frac{\partial^2}{\partial h^2} \right) V_{\mathrm{CW}} .
\end{equation}
The first-derivative term enforces the minimum condition for quantum corrections, which explicitly incorporates the tadpole condition into the derivation. By combining all of the aforementioned equations, we obtain the final expressions for the Higgs self-couplings at the next-to-leading order
\begin{equation}
\label{eq:lamuda3}
\lambda_{hhh}^{(1)} = \frac{3\left[M_h^2\right]_V}{v} + \frac{a_2^3 v^3}{4\pi^2 m_S^2} - \frac{3 y_t^4 v}{4\pi^2},
\end{equation}
\begin{equation}
\label{eq:lamuda4}
\lambda_{hhhh}^{(1)} = \frac{3\left[M_h^2\right]_V}{v^2} + \frac{3 a_2^3 v^2}{2\pi^2 m_S^2} - \frac{a_2^4 v^4}{2\pi^2 m_S^4} - \frac{3 y_t^4}{\pi^2}.
\end{equation}

%%%%%%%%%%%%%%%%%%%%%%%%%%%%%%%%%%%%%%%%%%%%%%%%
%				
%%%%%%%%%%%%%%%%%%%%%%%%%%%%%%%%%%%%%%%%%%%%%%%%

\section{Constraints from the Higgs Self-Couplings}
\label{sec:constraints}

In this section, we investigate the compatibility between the model parameter spaces quoted in Section~\ref{sec:phenomenology} and existing as well as prospective measurements of the Higgs self-couplings. We further incorporate the theoretical constraint from perturbative unitarity for the $HH \rightarrow HH$ scattering process into the analysis \cite{Stylianou_2024, Liu_2018, Chang_2020}. The zeroth partial-wave amplitude of this process is required to satisfy $a_{ii}^0 \le \frac{1}{2}$ across a sufficiently broad energy range. At the tree level, the amplitude takes the following form \cite{Stylianou_2024}
\begin{align}
	a_{ii}^{0}=\frac{3M_{h}^{2}\sqrt{s-4M_{h}^{2}}}{32\pi s(s-M_{h}^{2})v^{2}}\left[\kappa_{4}(s-M_{h}^{2})-3\kappa_{3}^{2}M_{h}^{2}+\frac{6\kappa_{3}^{2}M_{h}^{2}(s-M_{h}^{2})}{s-4M_{h}^{2}}\ln\left(\frac{s}{M_{h}^{2}} - 3\right)\right],
\end{align}
where $s$ corresponds to the squared center-of-mass energy.

The measurement of the Higgs boson self-coupling is one of the core scientific objectives of the LHC physics program. Corresponding searches have been performed by the ATLAS and CMS Collaborations across a wide range of final states, utilizing proton–proton collision datasets collected at $\sqrt{s}=13\,\mathrm{TeV}$ during the Large Hadron Collider Run 2 \cite{ATLAS:2019qdc,ATLAS:2022jtk,Arcadi:2024ukq,ATLAS:2023elc,ATLAS:2023gzn,ATLAS:2024pov,ATLAS:2024lhu,ATLAS:2024ish,ATLAS:2024xcs,ATLAS:2025hhd,CMS:2020tkr,CMS:2022cpr,CMS:2022gjd,CMS:2022omp,CMS:2024fkb,CMS:2024awa,CMS:2025ngq,CMS:2026nuu}. The latest combined analysis conducted by the ATLAS and CMS Collaborations yields a $95\%$ confidence-level constraint of $-0.71 < \kappa_3 < 6.1$ \cite{CMS:2026nuu}. Refined sensitivity projections for the HL-LHC, obtained from combined channel analyses with improved analysis methodologies based on 3000 $\mathrm{fb}^{-1}$ of integrated luminosity, give an expected measurement uncertainty of $30\%$ \cite{cepeda2019higgsphysicshllhchelhc,ATLAS:2025eii}. Future electron-positron colliders such as the ILC, CLIC, CEPC, and FCC-ee, characterized by clean collision environments and precisely calculable initial states, are capable of delivering extremely precise measurements of single Higgs boson production; the uncertainty on the Higgs trilinear coupling is estimated to fall in the range of $15\%$ to $40\%$. A percent-level determination of the Higgs self-coupling is projected to be achievable at future high-energy hadron collider facilities such as the FCC-hh \cite{Fcc-hh}. 

We conduct a numerical scan over the parameter spaces given by
\begin{equation}
	1\ \mathrm{GeV} < m_S < 100\ \mathrm{TeV}, \qquad -20 < a_2 < 20,
\end{equation}
where the SM input parameters employed in this analysis are taken from the latest dataset released by the Particle Data Group \cite{ParticleDataGroup:2026aaa}. 

\begin{figure}[htbp]
    \centering
    \begin{subfigure}[b]{0.49\textwidth}
        \centering
        \includegraphics[width=\textwidth]{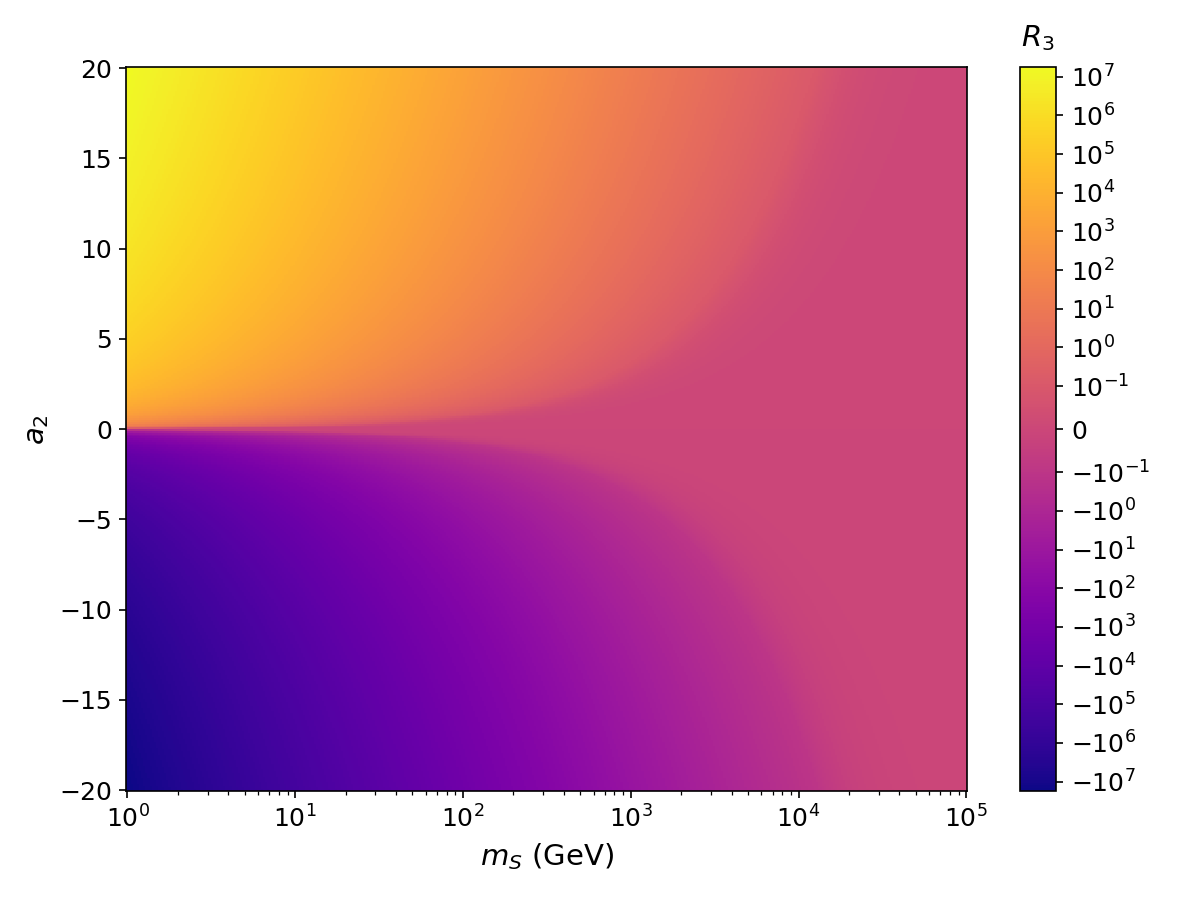}
        \caption{}
        \label{fig:R3-plane}
    \end{subfigure}
  \hfill
    \begin{subfigure}[b]{0.49\textwidth}
        \centering
        \includegraphics[width=\textwidth]{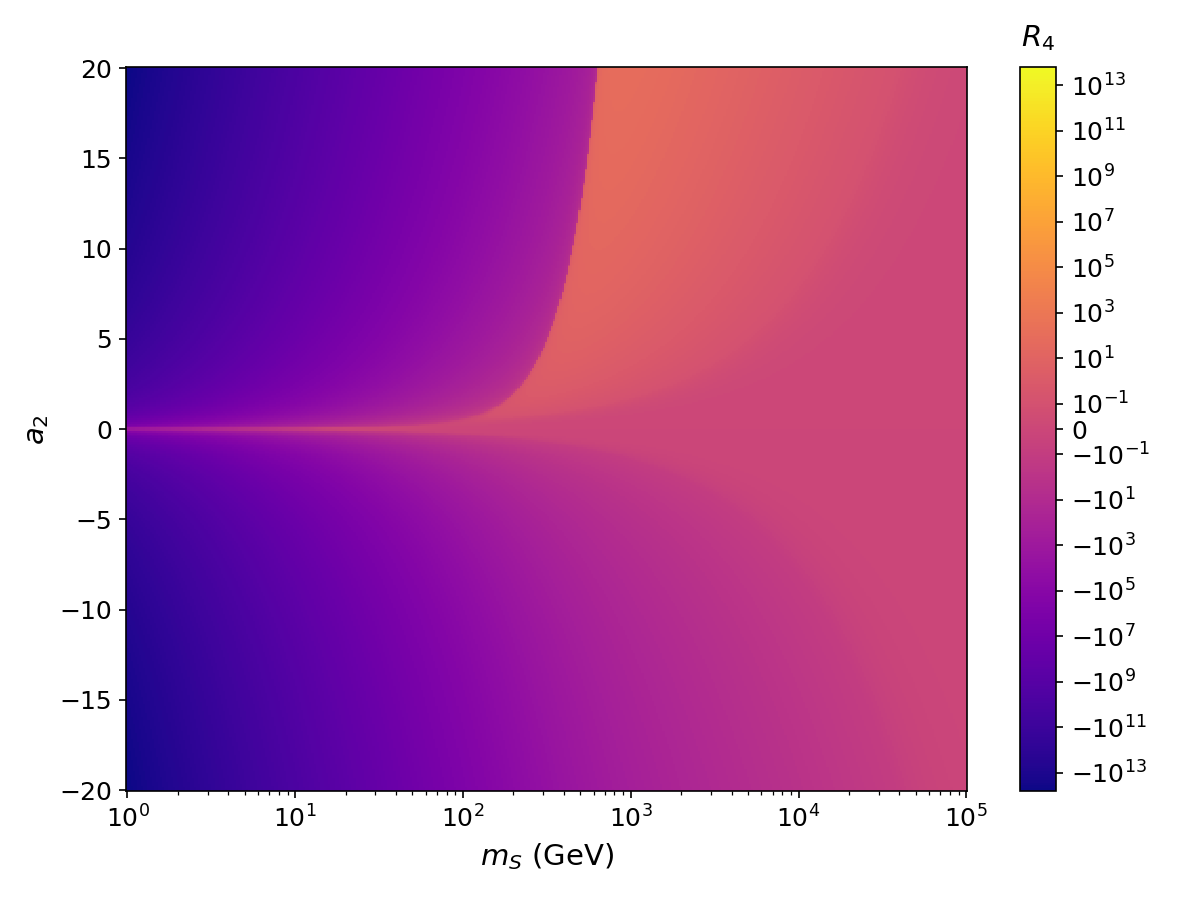}
        \caption{}
        \label{fig:R4-plane}
    \end{subfigure}
    \caption{\justifying\setlength{\baselineskip}{14pt}Distribution of the relative deviations $R_3$ (left panel) and $R_4$ (right panel) over the $(m_S,\; a_2)$ parameter space. The color gradient ranging from deep purple to light yellow corresponds to the variation of $R_3$ and $R_4$ values from negative to positive.}
    \label{fig:R3R4}
\end{figure}

Figure~\ref{fig:R3R4} presents the results for $R_3$ and $R_4$, which are defined as the relative deviations from the SM couplings
\begin{align}
	R_3 \equiv \frac{\lambda_{hhh}(\mathrm{Z}_2\mathrm{xSM}) - \lambda_{hhh}(\mathrm{SM})}{\lambda_{hhh}(\mathrm{SM})}=\kappa_3-1, \qquad
	R_4 \equiv \frac{\lambda_{hhhh}(\mathrm{Z}_2\mathrm{xSM}) - \lambda_{hhhh}(\mathrm{SM})}{\lambda_{hhhh}(\mathrm{SM})}=\kappa_4-1.
\end{align}
Across the $(m_S,\; a_2)$ plane, the color gradient ranges from deep purple to light yellow, which corresponds to the transition of $R_3$ and $R_4$ values from negative to positive. Our analysis reveals that at one-loop order, the absolute value of $R_3$ is symmetric about $a_2 = 0$, satisfying the relation $R_3(m_S,a_2)=-R_3(m_S,-a_2)$. Owing to the presence of the $-a_2^4$ term in Eq.~\eqref{eq:lamuda4}, $R_4$ takes negative values for small DM masses. Furthermore, both $R_3$ and $R_4$ exhibit a distinct decoupling effect: as the DM mass $m_S$ increases, their absolute values decrease, since both quantities scale as negative powers of $m_S$. In the perturbative limit of $a_2$, both quantities approach zero when $m_S>60\ \mathrm{TeV}$. Analogous decoupling phenomena have been observed and reported in multiple models beyond the Standard Model (BSM) \cite{Kanemura_2017, Braathen_2019, Braathen_2020, Braathen_2021}. This decoupling behavior indicates that the corrections to the Higgs self-couplings induced by BSM particle vanish as a power law when the mass of the particle becomes sufficiently large. 

We next investigate the constraints imposed by $\kappa_3$ and $\kappa_4$ upon the DM parameter space. As presented in Fig.~\ref{fig:contours_kappa}, an array of $\kappa_3$ contours is plotted alongside the perturbative unitarity exclusion boundary in the $(m_S,\; a_2)$ plane, with all parameter regions referenced in Section~\ref{sec:phenomenology} adopted from existing literature\footnote{\justifying\setlength{\baselineskip}{14pt}To facilitate a direct comparison of the constraint efficacy arising from the Higgs invisible decay channel and $\kappa_3$, the parameter spaces cited from prior work exclude constraints derived from Higgs invisible decay measurements; instead, the boundary corresponding to Higgs invisible decay constraints is represented as a separate independent curve.}. The green-shaded, blue-shaded, and orange-shaded regions correspond respectively to the WIMP-RM\cite{yu2025newconstraintssingletscalar}, WIMP-LM\cite{Escudero_Abenza_2025}, and SIMP\cite{Belanger:2026ctm} scenarios. The full viable parameter space associated with the freeze-in mechanism, as reported in Ref.~\cite{arcadi2024higgsportaldarkmatter} for reheating temperatures satisfying \(T_R \ge 0.2\,\mathrm{GeV}\), is identified as the FISC parameter space, bounded by the purple dotted contour on the plane. As the LO expressions provided in Eqs.~\eqref{dd} through \eqref{hid} scale as $a_2^2$, no existing experimental dataset can exclude the regime of $a_2 < 0$.\footnote{\justifying\setlength{\baselineskip}{14pt}Within the relic density calculation framework, only the $SS \rightarrow hh$ annihilation cross section contains odd powers of $a_2$ (see Eq.~(A4) of \cite{Cline_2013}). Nevertheless, this channel contributes at most one-third of the total relic density, and its associated effect is negligible for the analysis presented in this work.} For this parameter regime, the theoretical requirement of vacuum stability imposes the constraint $\lambda b_4/6 > a_2^2$ \cite{yu2025newconstraintssingletscalar}.

\begin{figure}[htbp]
	\centering
	\includegraphics[width=0.75\textwidth]{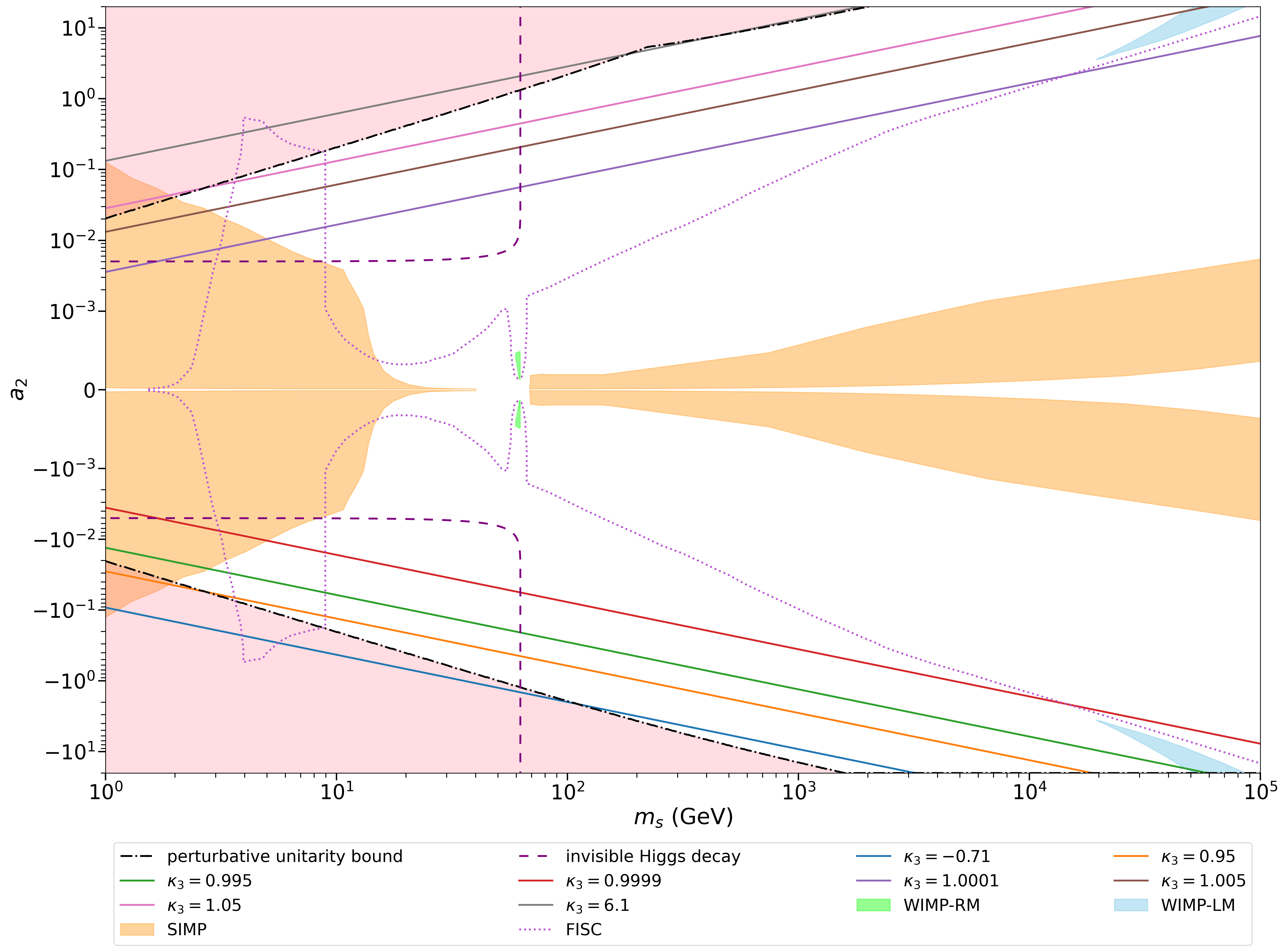}
	\caption{\justifying\setlength{\baselineskip}{14pt}The contours of $\kappa_3$ and the perturbative unitarity bound in the $(m_S,\; a_2)$ parameter plane. The green-shaded, blue-shaded, and orange-shaded regions represent the parameter spaces associated with the WIMP-RM\cite{yu2025newconstraintssingletscalar}, WIMP-LM\cite{Escudero_Abenza_2025} and SIMP\cite{Belanger:2026ctm} scenarios, respectively. The parameter domain of FISC scenario is demarcated by the enclosed purple dotted contour\cite{arcadi2024higgsportaldarkmatter}.}
	\label{fig:contours_kappa}
\end{figure}

The solid lines of different colors correspond to distinct values of $\kappa_3$. Owing to the positive correlation between $\kappa_3$ and $a_2$, contours of $\kappa_3$ with ascending values are ordered from bottom to top along the vertical axis. The region bounded by two adjacent contours corresponds to the parameter space where $\kappa_3$ falls between the corresponding lower and upper bounds. Along any given $\kappa_3$ contour, the magnitude of the portal coupling $|a_2|$ exhibits an increasing trend as the DM mass increases. This observation can be directly attributed to the decoupling effect: the correction to the Higgs self-coupling gradually diminishes as the DM mass increases. Consequently, a larger magnitude of portal coupling $|a_2|$ is required to accommodate the correction induced by a larger DM mass.

The black dot-dashed line denotes the exclusion boundary derived from perturbative unitarity, and the excluded parameter region is shaded in pink. Perturbative unitarity constraints, when formally applied, allow for the exclusion of a marginal portion of the low-mass regions in the SIMP and FISC parameter spaces, provided these regions have not been previously excluded by current measurements of the Higgs invisible decay channel. The tree-level zeroth partial wave amplitude $a_{ii}^0$ for $HH \rightarrow HH$ scattering becomes sufficiently small, such that the unitarity condition $a_{ii}^0 \leq 1/2$ is always satisfied across the entire high-mass region. As a consequence of the decoupling effect, the corrections to both the trilinear and quartic Higgs self-couplings approach zero when the DM mass exceeds 60 TeV in the perturbative limit of $a_2$. 

Existing experimental measurements of the trilinear Higgs self-coupling $\kappa_3$ currently lack sufficient precision to impose effective constraints on the DM parameter space. For illustrative context, the latest combined confidence interval for $\kappa_3$, released jointly by the ATLAS and CMS collaborations as $-0.71 < \kappa_3 < 6.1$, exhibits an excessively broad permitted range, such that nearly all parameter domains detailed in Section~\ref{sec:phenomenology} fall entirely within this interval. In the low DM mass regime, the stringency of this $\kappa_3$ constraint is even lower than the exclusion limit derived from perturbative unitarity, a constraint that itself is far less restrictive than the bounds obtained from contemporary measurements of the Higgs invisible decay channel.

The proposed FCC-hh collider is projected to achieve a measurement uncertainty of approximately $5\%$ for the Higgs trilinear coupling. Within the SM, this level of precision corresponds to the interval $0.95 < \kappa_3 < 1.05$ at $68\%$ CL \cite{Gon_alves_2018}. The yielded exclusion limit exhibits sensitivity level comparable to that derived from perturbative unitarity bound. Despite this comparatively narrow permitted range, the associated constraint is only capable of excluding a marginal fraction of the SIMP and FISC parameter space sectors tied to DM mass below 10 GeV—sectors that have already been ruled out by present-day measurements of the Higgs invisible decay channel. Conversely, if the FCC-hh measurement yields $\kappa_3 > 1.05$ or $\kappa_3< 0.95$, almost all of the investigated parameter space will be excluded.

For the extreme measurement uncertainties reduced to the order of $\mathcal{O}(0.1\%)$, the permissible range of $\kappa_3$ in the SM is restricted to the narrow interval $0.995 < \kappa_3 < 1.005$. The exclusion bound derived from this range covers only a limited subset of the SIMP and FISC parameter domains corresponding to DM masses below 10 GeV, as well as a marginal fraction of the WIMP-LM parameter space. Nevertheless, the sensitivity of this projected exclusion limit is markedly inferior to the existing experimental constraint from current ATLAS measurements, which sets an upper bound of $\mathrm{BR}_{\mathrm{inv}} \leq 0.107$ on the invisible decay branching fraction of the Higgs boson. It is worth noting that this specific subset of SIMP and FISC parameter domains for DM masses below 10 GeV has already been ruled out by existing constraints on Higgs invisible decays\cite{Belanger:2026ctm,arcadi2024higgsportaldarkmatter}. This projected exclusion limit only attains sensitivity levels comparable to the aforementioned existing experimental constraint when $\kappa_3$ lies within the narrow interval (0.9999, 1.0001).

\section{Conclusion}
\label{sec:conclusion}

Taking the real singlet scalar DM model ($\mathrm{Z}_2\mathrm{xSM}$) as a concrete research example, this work investigates the constraints imposed by Higgs self-couplings on Higgs portal DM models. In the $\overline{\mathrm{MS}}$ renormalization scheme, the effective trilinear and quartic Higgs couplings are derived from the one-loop effective potential of the Higgs sector. Contours of the trilinear coupling modifier $\kappa_3$ corresponding to different values, as well as the tree-level unitarity limit for $HH \rightarrow HH$ scattering, are plotted as a function of the DM mass $m_S$ and the portal coupling $a_2$ in Fig.~\ref{fig:contours_kappa}, from which constraints on DM properties can be directly extracted.

In the low DM mass regime, the constraining power afforded by the extant highest-precision measurement of $\kappa_3$ is markedly weaker than the exclusion limit derived from perturbative unitarity requirements, a bound that itself remains substantially less restrictive than the constraints extracted from contemporary measurements of the Higgs invisible decay channel. The constraining sensitivity attainable from $\kappa_3$ measurements reaches a level comparable to the current experimental upper bound on the Higgs invisible decay branching ratio $\mathrm{BR}_{\mathrm{inv}} \leq 0.107$, only under the condition that the measurement uncertainty of $\kappa_3$ is reduced to the $\mathcal{O}(0.01\%)$ scale, which maps to the allowed interval $0.9999 < \kappa_3 < 1.0001$. Both $\kappa_3$ and $\kappa_4$ manifest a clear decoupling behavior, with their magnitudes decreasing as the DM mass $m_S$ increases. Accordingly, the corrections to both couplings asymptotically approach zero when the DM mass exceeds 60 TeV, within the perturbative limit of the parameter $a_2$.

Contemporary LHC measurements of the Higgs trilinear self-coupling $\kappa_3$ lack the requisite precision to impose incremental constraints on the singlet scalar DM parameter space that extend beyond existing bounds derived from thermal relic density requirements, direct detection experiments, indirect detection experiments, and searches for invisible Higgs decays. For the planned future circular hadron collider (FCC-hh), scenarios in which the measurement uncertainty of $\kappa_3$ reaches the $\mathcal{O}(1\%)$ scale — corresponding to the allowed interval $0.95 < \kappa_3 < 1.05$ — yield excluded low-mass SIMP and FISC parameter regions that lie entirely within the exclusion domain established by invisible Higgs decay search analyses. At an extreme precision benchmark of $\mathcal{O}(0.1\%)$ measurement uncertainty, only a marginal fraction of the WIMP large mass (WIMP-LM) parameter space would be excluded. Such a precision level is, however, practically unachievable in realistic experimental implementations.

%%%%%%%%%%%%%%%%%%%%%%%%%%%%%%%%%%%%%%%%%%%%%%%%
%					
%%%%%%%%%%%%%%%%%%%%%%%%%%%%%%%%%%%%%%%%%%%%%%%%

\bibliography{higgs_coupling}

\end{document}